\documentclass[12pt,preprint]{aastex}

\pdfoutput=1

\begin{document}

\title{HI in HO \\ {\small Hoag's Object revisited}}

\author{Noah Brosch and Ido Finkelman\altaffilmark{}}
\affil{The Wise Observatory and the Raymond and  Beverly Sackler School of Physics and
Astronomy, the Faculty of Exact
Sciences, \\ Tel Aviv University, Tel Aviv 69978, Israel}
\email{noah@wise.tau.ac.il, idofinkelman@gmail.com}

\author{Tom Oosterloo\altaffilmark{}}
 \affil{Netherlands Institute for Radio Astronomy, ASTRON PO Box 2, 7990 AA Dwingeloo, \\ and Kapteyn Astronomical Institute, University of Groningen, P.O. Box 800, 9700 AV Groningen, the Netherlands}
\email{oosterloo@astron.nl}

\author{Gyula Jozsa\altaffilmark{}}
\affil{Netherlands Institute for Radio Astronomy, ASTRON PO Box 2, 7990 AA Dwingeloo, the Netherlands, and Argelander-Institut f\"ur Astronomie, Auf dem H\"ugel 71, D-53121 Bonn, Germany}
\email{jozsa@astron.nl}

\author{Alexei Moiseev\altaffilmark{}}
\affil{Special Astrophysical Observatory, Russian Academy of Sciences, Nizhniy Arkhyz, Karachai-Cherkessian Republic 357147, \\ and Sternberg Astronomical Institute of Lomonosov Moscow State University, Moscow 119992, Russia}
\email{moisav@sao.ru}


\begin{abstract}
We present new HI observations of Hoag's Object obtained with the Westerbork Synthesis Radio Telescope (WSRT). The data show that the luminous optical ring around the elliptical body has a bright HI counterpart that shares the kinematical properties of the optical ring. The entire HI structure is twice as large as the optical ring and shows a mild warp in its outer regions relative to the inner ring. We detect two additional HI sources close in redshift to that of Hoag's Object, and report on a newly identified SDSS optical companion galaxy. The HI sources are $\sim$0.3 and $\sim$1 Mpc away in projected distance, and the companion galaxy is also $\sim$1 Mpc away. Our main conclusion is that the HI detected in Hoag's Object shows no indication that this galaxy has experienced a recent (less than $\sim$1 Gyr ago) accretion event. At least one of the two additional HI detected objects does not have an optical counterpart. One possibility is that this object is an HI filament left over from an interaction shaping Hoag's Object, in which case this interaction must also  have occurred at least 1-2 Gyr ago.

\end{abstract}



\section{Introduction}
Hoag's Object (HO) is an intriguing galaxy. Discovered by Hoag (1950), it was the topic of  few research papers in the last six decades despite being the most perfect ring galaxy known today. While Hoag suggested that this could be a case of gravitational lensing since the ring seemed perfect on the 24-inch Jewett Schmidt telescope plates he examined, O'Connell et al. (1974) ruled out the lensing hypothesis by arguing that, in this case, the central object should have a mass-to-light ratio of 1500 M$_{\odot}$/L$_{\odot}$.

Brosch (1985) studied optical images of HO obtained at the Wise Observatory and HI synthesis observations with WSRT. The optical data showed that the central object has a de Vaucouleurs (r$^{1/4}$) profile and that the ring's surface brightness was significantly brighter than the extrapolated brightness of the central object at the same radius. The 12-hour synthesis HI observations were unsuccessful, yielding a 3$\sigma$ upper limit of 2.9 mJy over 33 km s$^{-1}$.
  Thus only an upper limit of 2.3$\times10^9$ M$_{\odot}$ could be set for the total HI content for an unresolved HI object at the position of Hoag's Object. Brosch proposed that HO was the result of star formation induced in a gaseous ring around an object similar to an elliptical galaxy that had a bar, which almost relaxed back into the central object.

Schweizer et al. (1987) presented optical observations with the Palomar 5-m telescope and with the Arecibo radio telescope. Their optical spectroscopy ruled out the gravitational lensing suggestion by showing that the ring and the central object are at the same redshift. The 21-cm Arecibo observations showed a two-horned profile with a width of 239 km s$^{-1}$ and a flux integral of 1.15$\pm$0.10 Jy km s$^{-1}$ centered on the optical redshift, from which Schweizer et al. concluded that Hoag's Object contained (7.0$\pm$0.1)$\times10^9$ M$_{\odot}$ of HI (converted to our assumed distance of 175.5 Mpc using H$_0$=73 km s$^{-1}$ Mpc$^{-1}$). They proposed that HO was the result of a major accretion event at least 2-3 Gyr in the past


Finkelman et al. (2011) analyzed HST archival observations and optical spectroscopy, including two-dimensional scanning Fabry-Perot interferograms of the H$\alpha$ line, obtained at the Russian Academy of Sciences 6-m BTA telescope. The HST images confirmed that the inner body has indeed a de Vaucouleurs surface brightness profile. The central object has the appearance of a slightly triaxial elliptical galaxy, and the spectroscopy showed that it is a fast rotator.

Finkelman et al. (2011) found that the ionized gas kinematics (in the ring) could be fitted with a circular rotation model, with its equatorial plane inclined relative to the line of sight by 18$^{\circ} \pm 4^{\circ}$. The ring was restricted to radii 14$\leq$r$\leq$28 arcsec and various fitting attempts indicated that  non-circular motion is present there in addition to the circular motion.
The ring showed a braided quasi-spiral structure of HII regions. Population synthesis of observed optical  spectra showed that the core object is older than 10 Gyr while the ring's stellar population is $\sim$1 Gyr, sustaining a low level of star formation at a rate of $\sim0.7$ M$_{\odot}$ yr$^{-1}$. The age of the stellar population in the ring was derived from a luminosity-weighted single stellar population, thus not excluding the presence of stars significantly older than 1 Gyr. Finkelman et al. (2011) proposed that a large HI mass was accreted soon after the formation of the elliptical core. The HI settled into a disk which now shows star formation triggered by the mild triaxiality of the core. Given the current star formation rate and the Are4cibo-detected HI content, Hoag's Object could remain in a quasi-steady-state for a Hubble time.

HI observations provide a complementary method to optical observations at revealing past interactions and possibly accretion events, since at very large radii, where HI is often found but the ionized gas component is weak or absent in the optical range, the kinematic timescales are long. Nevertheless, relatively little effort has been spent on studying the gas properties of Hoag's Object, with the exceptions of Brosch (1985) and Schweizer et al. (1987) mentioned above.


While the optical appearance of HO seems now reasonably well understood, the questions are where in relation to the luminous galaxy is the HI detected by Schweizer et al. (1987), how is this HI distributed, and what are its kinematic properties.   The presence or absence of morphological or kinematical irregularities in the HI distribution can provide essential inputs for evolution models of this object thus determining its place among other galaxies. To understand HO better we performed synthesis HI observations at the WSRT, expecting that these would enable the study of both the neutral gas distribution and of the HI kinematics.   The optical ring of Hoag's Object has a diameter of about 40 arcsec, thus the WSRT is well suited to spatially resolve the ring while still maintaining good column density sensitivity. We show here that the HI is arranged in a ring significantly larger than the optical one, is slightly warped at its outer regions, and contains at least 6.2$\times10^9$ M$_{\odot}$ of HI.

The structure of this paper is as follows. In Section~\ref{sec.obs} we describe the observations and their reductions. Section~\ref{sec.proc} details the reduction process, which had to be somewhat different from the standard WSRT reductions to reveal faint HI details near HO. The results are detailed in the same section, and we present in Section~\ref{sec.discuss} a discussion where HO is discussed in the context of other ringed galaxies. Section~\ref{sec.summary} summarizes this paper.

\section {Observations and data reduction}
\label{sec.obs}

Hoag's Object was observed for 4$\times$12+6 hours (54 hours total) with the Westerbork Synthesis Radio Telescope (WSRT) in the maxi-short configuration, with a 36m shortest baseline. 
The setup used a 20 MHz band centered on 12736 km s$^{-1}$ split into 1024 spectral channels, with the two orthogonal polarizations  averaged.

The data were calibrated in the standard way, using the Miriad package (Sault, Teuben \& Wright 1995). We used the data to make two data cubes, one using standard weighting (robust=0.4) to image the kinematics of the HI, and another cube made with natural weighting to have maximal sensitivity for detecting faint HI structures in the vicinity of HO. The velocity resolution of these data cubes is 8.8 km s$^{-1}$. The beam of the standard cube is 16$\times$45 arcsec (HPBW) and this cube has a 5$\sigma$ column density detection limit of 2.6$\times$10$^{19}$ cm$^{-2}$ over 20 km s$^{-1}$ and a mass detection limit of 1.2$\times10^8$ M$_{\odot}$ over 20 km s$^{-1}$. The 5$\sigma$ column density detection limit in the naturally weighted cube is  1.4$\times10^{19}$ cm$^{-2}$ over 20 km s$^{-1}$. This sensitivity is one order of magnitude better than that of the 1982 WSRT observations (Brosch 1985).



\section {Analysis and results}
\label{sec.proc}

\begin{figure}[t]
\centering{
  \includegraphics[width=15cm]{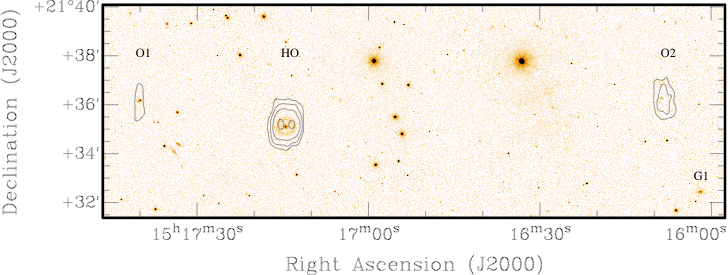}
  \caption{The field of Hoag's Object from the SDSS $r$-band image, with superposed HI contours and various objects labelled. The contour levels are (3, 6, 12 and 24)$\times10^{19}$ cm$^{-2}$. The companion HI cloud O1 is to the left of HO and the second HI companion, O2, is near the right edge of the plot. More detailed plots for each of the sources are shown in Figures 2 (HO), 5 (O1) and 6 (O2). A newly identified neighbor galaxy (see below) is marked G1 at the lower-right corner of the image.}
  \label{fig:all}}
\end{figure}
\begin{figure}[t]
\centering{
  \includegraphics[width=10cm]{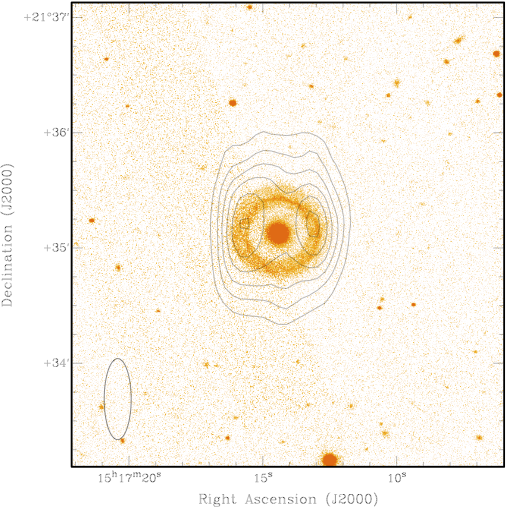}
  \includegraphics[width=10cm]{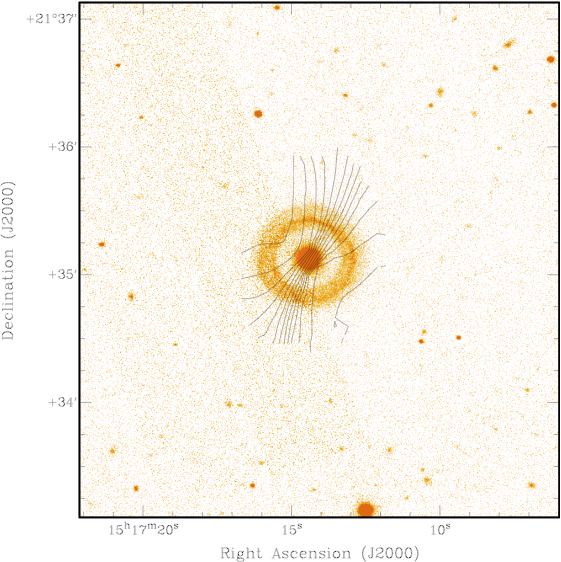}
  \caption{Hoag's object in HI. {\it Top panel:} The HI contours overplotted on the SDSS $r$-band image. The synthesized beam is plotted at the lower left corner. The contours plotted are (5, 10, 15, 20, 25 and 30)$\times 10^{19}$ cm$^{-2}$. {\it Bottom panel:} The velocity field of Hoag's object. The lowest contour is at 12670 km s$^{-1}$ and the step between consecutive contours is 10 km s$^{-1}$. The lowest velocity is at the lower-right side of the plot.}
  \label{fig:hoag}}
\end{figure}

Figure~\ref{fig:all} shows the optical image from SDSS with superposed HI contours from the standard-weighting map, detailing the detection of HI in HO and in two objects in its vicinity.  The figure shows that the HI is well detected in HO. The results for Hoag's Object are shown in better detail in Figure~\ref{fig:hoag}, where the top panel shows a SDSS image with the HI contours overlaid. The synthesized beam is shown at the lower left corner of this figure. The derived HO flux integral is 0.86 Jy km s$^{-1}$ which, for a distance of 175.5 Mpc, gives M(HI)=6.2$\times10^9$ M$_{\odot}$ with an uncertainty of 10\%. It is clear that the HI coincides with the optical ring and that no HI is detected inside it.


 The  HI distribution appears to have two maxima, East and West of the optical center. This is an artifact caused by the beam which is elongated in the N-S direction, since the observations were made with a linear antenna array oriented in the E-W direction.  A model of a uniform ring convolved with the synthesized
elongated beam yields exactly the observed structure. The HI distribution is definitely more extended than the optical image of the ring, perhaps twice as wide. On the other hand, the HI map does not show tails or extensions, as could have been expected were the HI recently acquired via a tidal interaction.

The lower panel of Figure~\ref{fig:hoag} shows the velocity field (VF) overplotted on the HO optical image. The VF was produced by fitting Hermite Gaussians to those profiles where the HI column density was above 5$\times10^{19}$ cm$^{-2}$. The velocity field appears quite regular, with no obvious disturbances, and is another argument against a recent interaction being the source of the HI.

Using the parameters of the kinematics of  the optical data from Finkelman et al. (2011), we constructed HI model cubes for Hoag's Object using  the Tilted Ring Fitting Code (TiRiFiC; Jozsa et al. 2007). Basically, the first model is a single ring of the same size as used in the modelling of the ionized gas velocity field
(radii 14-28 arcsec) of inclination $\sim$18$^{\circ}$ and using the same rotation velocities as Finkelman et al. The top panel of Figure 3 shows a position-velocity plot (not aligned with the major axis but along constant declination) of the model and the data. The figure shows that the model reproduces the data fairly well, except for the faint extensions at the most eastern and western sides of the ring, at velocities away from the systemic velocity.

\begin{figure}[t]
\centering{
   \includegraphics[width=10cm]{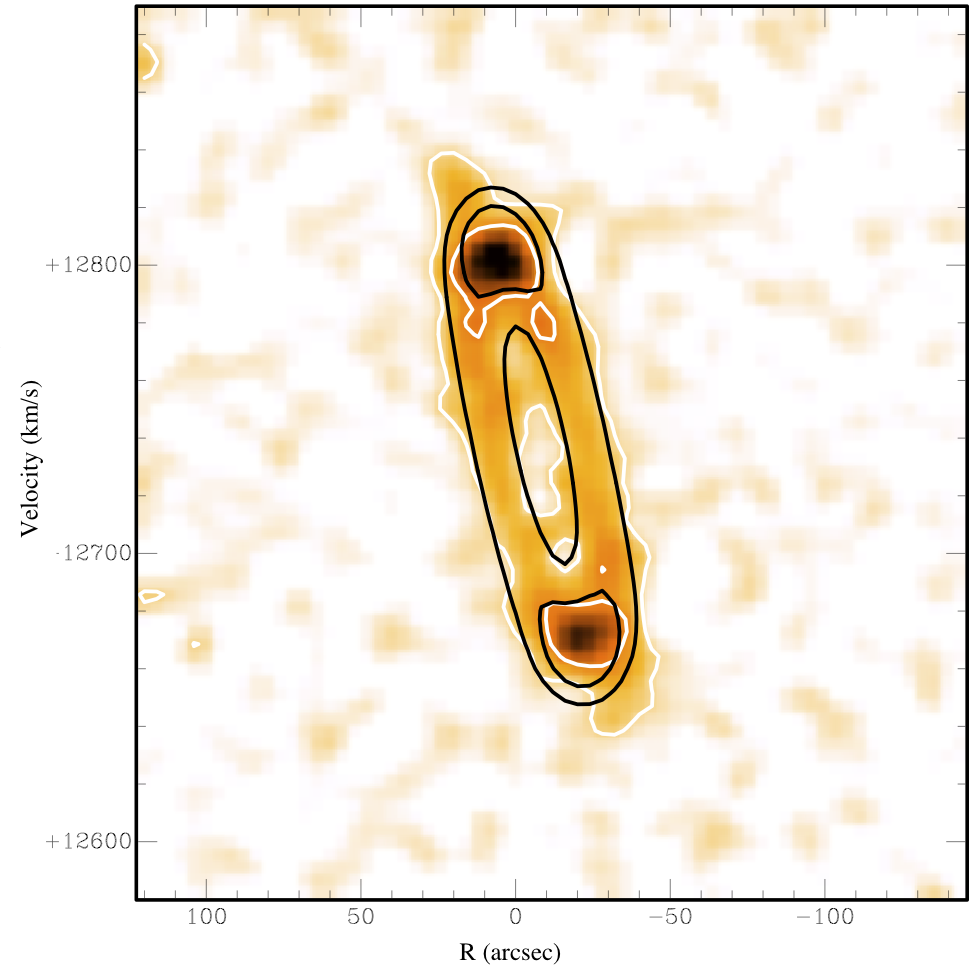}
  \includegraphics[width=10cm]{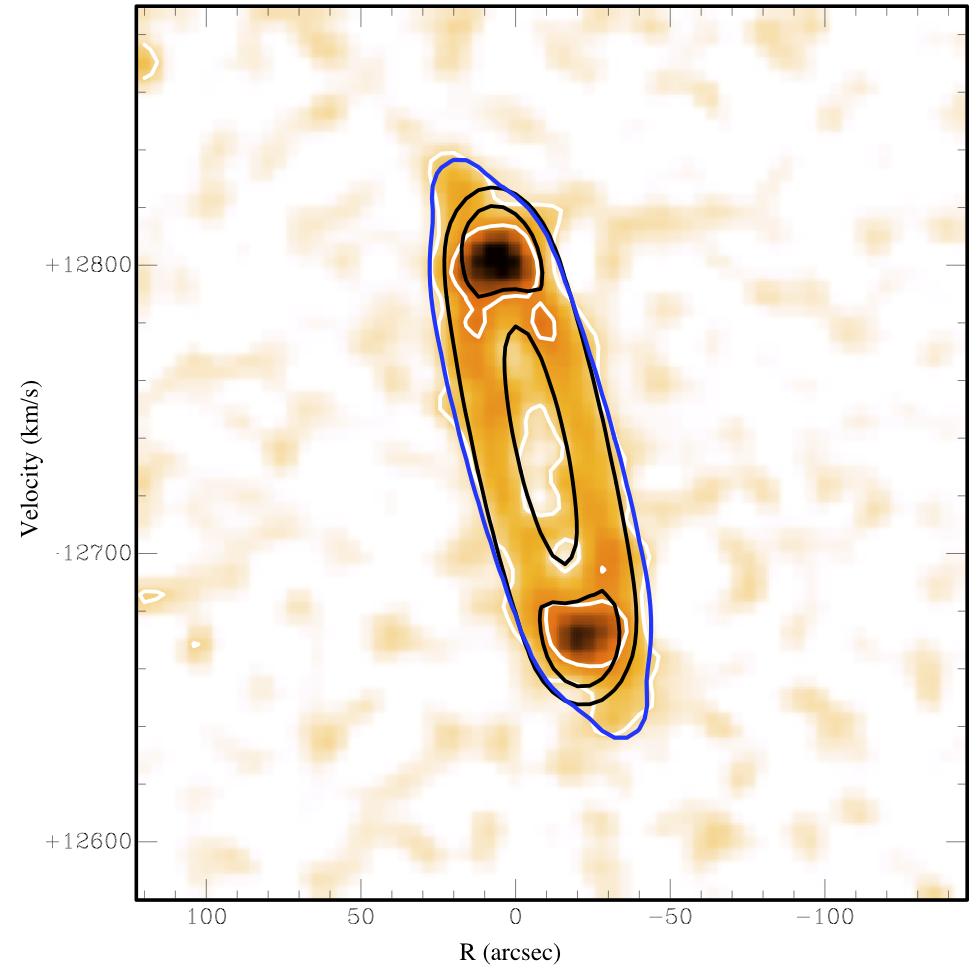}
  \caption{Models fitted to the HI distribution in Hoag's Object. {\it Top panel:} Single-ring fit, with the equatorial plane of the ring tilted 18$^{\circ}$ with respect to the line of sight. {\it Bottom panel:} Two-ring fit with the outer ring tilted by 7$^{\circ}$ with respect to the inner ring and with the same rotational velocity.}
  \label{fig:models}}
\end{figure}

These HI extensions suggest that the HI ring extends  beyond the optical ring, but at lower column densities compared to the inner  HI. We modelled this by extending the previous model with a second ring covering radii between 28 and 42 arcsec. Given the approximately face-on orientation of the HI in HO and the limited spatial resolution, it is difficult to obtain a sensible result using a free fit of all parameters. Therefore, we assumed that the rotation curve of HO is flat out to the outermost radii. Given that the outer HI has larger projected rotation velocities, this means that the outer ring has to have a slightly higher inclination than the inner HI.

We obtain a satisfactory model by using an inclination for the outer HI ring of 25$^{\circ}$, or 7$^{\circ}$ more than that of the inner ring.  The bottom panel of Figure 3 shows the position-velocity plot of this model, illustrating that this model describes the data quite well. The small inclination increase with radius implies that the HI ring in HO is slightly warped at large radius. Such warps are a common feature in any galaxy with an HI disk extending beyond the optical disk, including early-type galaxies with HI disks. Figure 4 summarizes the parameters of our final kinematical model, implying that in HO we see a stellar ring with a diameter of $\sim$47 kpc embedded in and surrounded by a $\sim$71 kpc wide HI ring.

\begin{figure}[t]
\centering{
  \includegraphics[width=12cm]{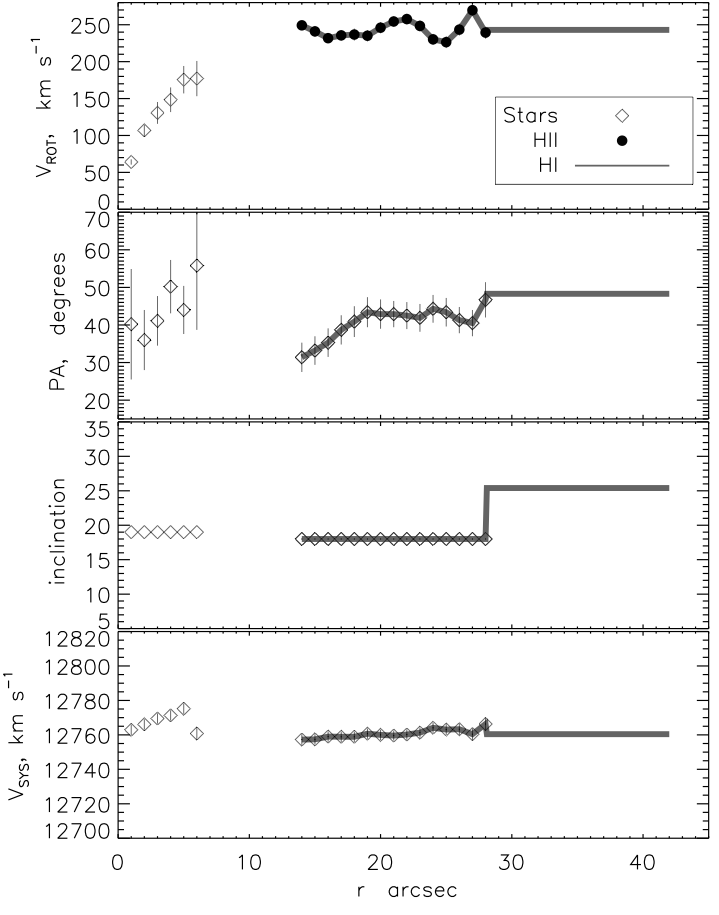}
  \caption{Joint optical-HI plot describing the variation in the kinematical properties of Hoag's Object. The parameters shown vs. radius are  rotational velocity (top), position angle (second from top), inclination (second from bottom), and systemic velocity (bottom).}
  \label{fig:models2}}
\end{figure}

The results discussed above are based on an HI cube made with standard weighting. We also made a naturally weighted cube to search for faint emission around HO. In this data cube we detect two additional HI sources, very faint, near the redshift of HO, that are also plotted in Figure 1. A few more galaxies are detected in the data, but these have systemic velocities about 1000 km s$^{-1}$ lower and are most likely not related, thus have not been plotted in the figure. The two sources possibly associated with HO are called here Object 1 (O1) and Object 2 (O2). The parameters of these companion objects are given in Table 1. These are very faint detections, not visible in the data cube with standard weighting because the signal is diluted by a large line width (see Fig. 3). However, because of their extent, they contain considerable amounts of HI.



\begin{table}
\label{t.objects}
\begin{footnotesize}
\begin{center}
\vspace{0.5cm}
\begin{tabular}{|c|c|c|c|c|}
\hline
Object/Parameter & HO & O1 & O2 & G1 \\
\hline
$\alpha$ J2000            & 15:17:14 & 15:17:40  & 15:16:09 & 15:16:01 \\
$\delta$ J2000           & +21:35:08  & +21:36:19 & +21:36:10 & +21:32:27\\
cz$_{\odot}$ [km s$^{-1}$] & 12736      & 12700     & 12629    & 12676 \\
M(HI) [10$^9$ M$_{\odot}$] & 6.2        & 0.39      & 1.2      & - \\
\hline
\end{tabular}
\caption{Objects in the HO complex}
\end{center}
Note to Table 1: Object G1 is the companion galaxy mentioned in the text and marked in Fig. 1, which does not have detectable HI.
\end{footnotesize}
\end{table}

SDSS $r$-band images with overplotted HI contours for O1 and O2 are shown in Figures 5 and 6. Figure 5 shows the total HI of O1 in the top panel and a position-velocity plot at constant declination in the bottom panel. Figure 6 shows O2, for which the HI signal is too faint to produce a useful position-velocity plot. Both objects seem to have faint optical counterparts within the synthesized beams, as Figures 5 and 6 show. SDSS reports an object at each of the locations and near O1 there is also a star.



\begin{figure}[t]
\centering{
  \includegraphics[width=10cm]{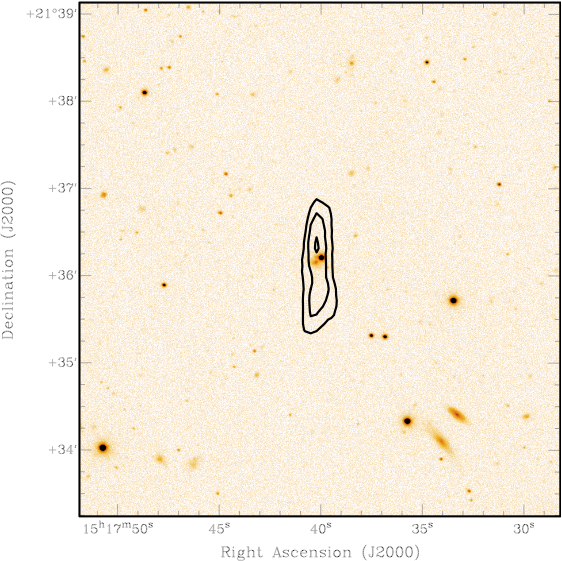}
  \includegraphics[width=10cm]{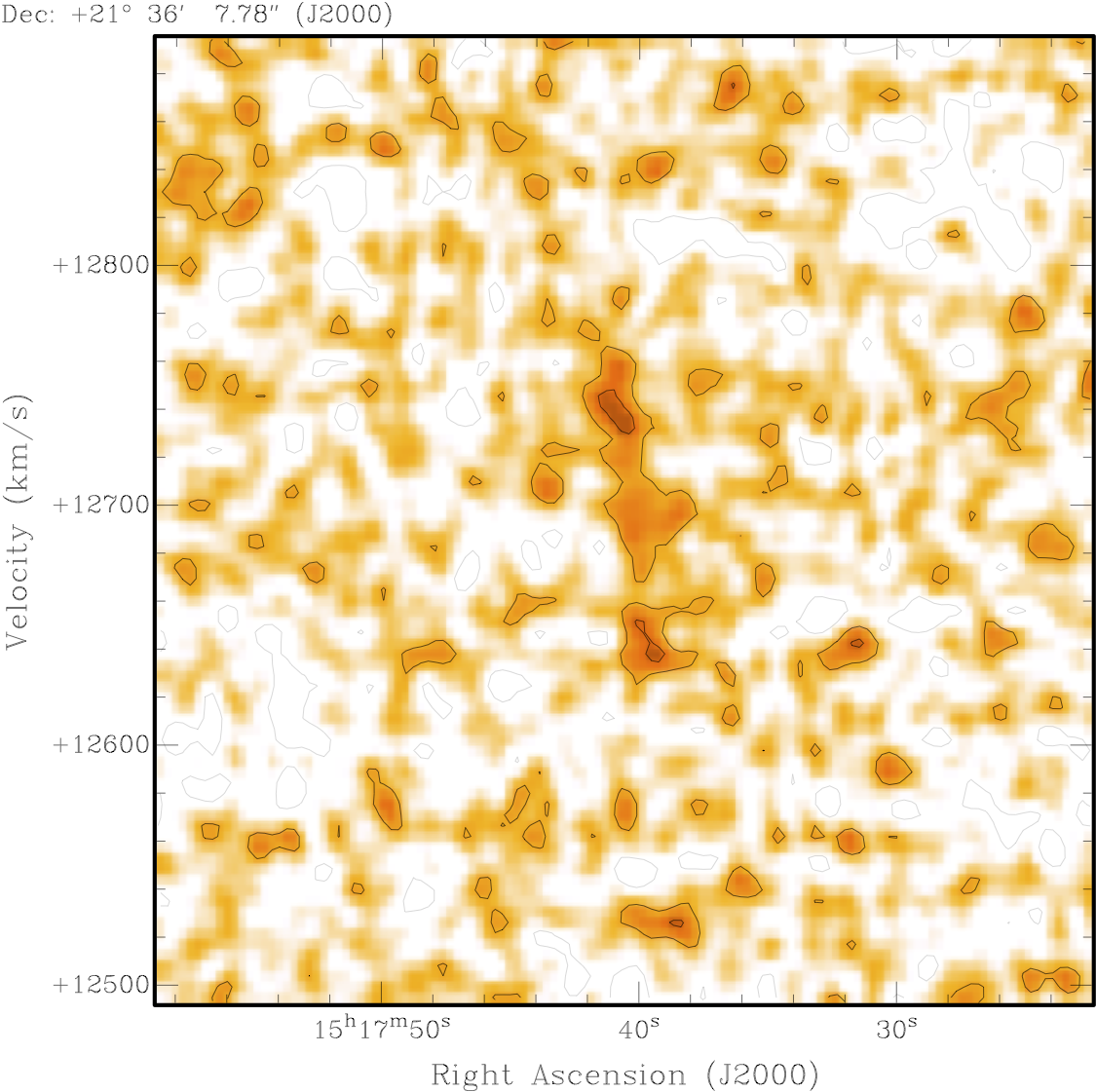}
  \caption{HI companion of Hoag's Object, called here O1 and detected using natural weighting of the WSRT channels. {\it Top panel:} The HI contours of O1 overplotted on the SDSS $r$-band image. The contours plotted are (3, 4 and 5)$\times 10^{19}$ H-atoms cm$^{-2}$. {\it Bottom panel:} A velocity-right ascension plot of O1. The contours plotted are -0.33, 0.33, 0.66 mJy beam$^{-1}$.}
  \label{fig:O1}}
\end{figure}

\begin{figure}[t]
\centering{
  \includegraphics[width=12cm]{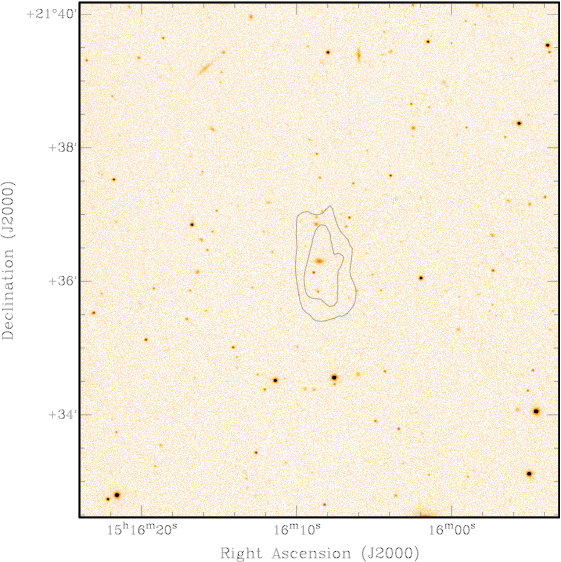}
  \caption{HI companion of Hoag's Object, called here O2 and detected in the same manner as O1.  Contours on the levels of (3 and 6)$\times 10^{19}$ H-atoms cm$^{-2}$ plotted on the SDSS $r$-band image.}
  \label{fig:O2}}
\end{figure}


The diffuse object in the immediate vicinity of O1 and adjacent to the bright star is the galaxy SDSS J151740.21+213609.7 which has a well-established spectroscopic redshift of z=0.143, implying that it is a background object not related to HO. The diffuse object near the center of O2 is the galaxy SDSS J151608.49+213618.4, which lacks a redshift. However, its apparent magnitude and colors, $r$=18.65 and ($g-r$)= 0.34 with ($r-i$)=0.22, put this object in the ``green valley'' location, if at the distance of Hoag's Object. Other SDSS objects near the HI locations of O1 or O2 also do not have optical spectral information but have bluish colors and the appearance of dwarf galaxies.

\section {Discussion}
\label{sec.discuss}

We mentioned in the introduction that Schweizer et al. (1987) measured a flux integral of 1.15 Jy km s$^{-1}$ with Arecibo, whereas we measure only 0.86 Jy km s$^{-1}$ from the synthesized map. These values are consistent, given their errors, and since the WSRT configuration, with a smallest baseline of 36m, did not miss much widespread HI. 

Finkelman et al. (2011) found no observational evidence to support late merging events in the evolution of this galaxy; the main stellar body was found to be at least 10 Gyr old and deep optical images from the 6-m BTA telescope did not reveal faint tidal tails or low surface brightness features around the galaxy, up to projected distance $\sim$150 kpc. The main result of the present paper is that, in the WSRT data shown here, we do not see any {\sl direct} evidence for an accretion event. The HI does extend somewhat beyond the optical ring, but the kinematics are regular. The orbital time at the outer edge of the optical ring is $\sim$6.7$\times 10^8$ yr and this is of order the crossing time $\tau$; since the relaxation time for a gas system is a few $\tau$, the lack of disturbances in the HI suggests that HO has not experienced a major accretion event in the last 1-2 Gyr.

Finkelman et al. (2011) argued that, since the nearest neighbor to Hoag's Object with an optically-measured redshift was about 3 Mpc away, HO must be a relatively isolated galaxy. A repeated search of the HO neighborhood using NED revealed a closer companion: 2MASX J15160166+2132270=SDSS J151601.65+213226.6, called here G1, with an SDSS DR9 redshift of 12676 km s$^{-1}$. The galaxy is visible near the lower-right corner of Figure 1 as a fuzzy elliptical blob. Inspecting the SDSS DR9 data shows an Sa shape and an absorption line spectrum. The SDSS absolute magnitude M$_r$=–-20.33 and color ($r-i$)=+0.87 puts this object in the ``red sequence`` location of the color-magnitude plot for galaxies.  This is also supported by the SDSS spectrum which does not show emission lines that could be associated with star formation. The newly reported optical companion is 17.2 arcmin=877 kpc projected distance and 60 km s$^{-1}$ away from HO, and only $\sim$100 kpc and 42 km s$^{-1}$ from the O2 HI cloud.

The two HI companions reported here, O1 and O2, are at $\sim$300 kpc and $\sim$1 Mpc projected distances respectively from HO. It is clear that all four bodies form a single complex, although no signs of ongoing mutual interactions are visible. It is interesting that O1 does not have an optical counterpart, despite having a reasonably wide HI line ($\sim$140 km s$^{-1}$, see bottom panel of Fig. 5). Assuming a simple Tully-Fisher relation, such a line width would imply an absolute magnitude of about --17, thus an apparent magnitude of $\sim$19, only slightly fainter than SDSS J151740.21+213609.7 that could mask its presence.

The optical counterpart of O1 could be a hidden dwarf galaxy, or this may mean that the HI in O1 could be the remnant of some galaxy-galaxy interaction, possibly one involving HO. Given its separation from HO, and assuming relative velocities of 200 km s$^{-1}$, such an interaction could have occurred not more recent than $\sim$1-2 Gyr ago.

We have shown in Figure 1 that HO and its two HI cloud companions are located on an approximately traight linear structure. The optical companion galaxy G1, O1 and HO lie actually on an even straighter line. This situation, of some optical galaxies and HI clouds being arranged in a linear shape is reminiscent of the finding of Beygu et al. (2013). They argue that a galaxy  triplet is growing along a filament in a void in the galaxy distribution from HI it concentrates out of the IGM.  Other similar linear structures were detected by Zitrin \& Brosch (2008).

Regarding the morphological peculiarity of Hoag's Object, the observations reported here and in Finkelman et al. (2011) support an interpretation that HO is another ringed galaxy where, by chance, we observe the gas-and-stars ring approximately pole-on, orbiting in or near the equatorial plane of the central body with its outer parts being slightly warped. 
 Such a configuration, where an outer star-forming ring surrounds an early-type galaxy, is reminiscent of the outer UV disks observed around lenticular galaxies by e.g. Ilyina \& Sil'chenko (2011) that sometimes appear as UV-bright rings (XUV rings). In fact, ESO 381-47, which is an S0 galaxy with a faint 30-kpc wide stellar ring and a much wider 90-kpc HI ring, seems to be very similar to Hoag's Object but at 61.2 Mpc is much closer (Donovan et al. 2009). The difference, in this case, is that while the objects discussed by Ilyina \& Sil'chenko and by Donovan et al. are early-type disks, Finkelman et al. (2011)  showed that the central object of HO matches a triaxial elliptical classification. 

  In this context, we mention that very large, regular HI structures are sometimes encountered around fairly isolated early-type galaxies (Oosterloo et al. 2010; Serra et al. 2012), and many of these structures are warped. 
   It is clear that Hoag's Object belongs to the class D galaxies, ``where most of the H I is found in a fairly regularly rotating disc or ring'' (Oosterloo et al.), best shown by NGC 3945 of morphological type (R)SB0, and NGC 5582 (type E) in Serra et al. (2012).

This paper adds one more object to the set of early-type galaxies with outer rings of stars and HI. Here too, as in the other cases, the origin of the gas is not clear-cut. This could be an early accretion event, $\sim$10 Gyr ago as suggested by Finkelman et al. (2011) but not later than a few Gyr ago. Or this could be on-going galaxy formation with gas accreted from the cosmic web that is manifested not only as HO's ring, but also as the O1 and O2 objects, although no observational evidence for that exists at present.

\section{Summary}
\label{sec.summary}

We presented HI synthesis observations of Hoag's Object obtained with the WSRT. These show that HO has an HI ring containing $\sim6 \times 10^9$ M$_{\odot}$ of HI that extends beyond the optical ring of this galaxy. The HI shares the kinematics of the optical ring. Outside the optical ring, the HI ring shows a slight warp. The kinematics of the HI are very regular, and there is no indication that Hoag's Object experienced a recent accretion event. From the kinematics of the HI we conclude that any accretion event could have happened no later than 1-2 Gyr ago.

We identified an optical companion that is an early-type disk $\sim$1 Mpc away in projected distance, and two 21cm clouds,  one at $\sim$300 kpc and containing $\sim4 \times 10^8$ M$_{\odot}$ of HI and the other more distant ($\sim$1 Mpc) with $\sim10^9$ M$_{\odot}$ of HI, at least one of the clouds lacking a confirmed optical counterpart. All four bodies may be part of a $\sim$linear structure extending over $\sim$1.5 Mpc, in the outskirts of a loose grouping of some 60 galaxies about 3-6 Mpc away.


\section{Acknowledgements}
AM is grateful for the financial support of the `Dynasty Foundation' and RAS programme  OFN-17. We acknowledge the allocation of observing time for this project by the WSRT. The Westerbork Synthesis Radio Telescope is operated by the ASTRON (Netherlands Institute for Radio Astronomy) with support from the Netherlands Foundation for Scientific Research (NWO). This research has made use of the NASA/IPAC Extragalactic Database (NED), which is operated by the Jet Propulsion Laboratory, California Institute of Technology, under contract with NASA.

Funding for the SDSS and SDSS-II has been provided by the Alfred P. Sloan Foundation, the Participating Institutions, the National Science Foundation, the U.S. Department of Energy, the National Aeronautics and Space Administration, the Japanese Monbukagakusho, the Max Planck Society, and the Higher Education Funding Council for England. The SDSS Web Site is http://www.sdss.org/.

The SDSS is managed by the Astrophysical Research Consortium for the Participating Institutions. The Participating Institutions are the American Museum of Natural History, Astrophysical Institute Potsdam, University of Basel, University of Cambridge, Case Western Reserve University, University of Chicago, Drexel University, Fermilab, the Institute for Advanced Study, the Japan Participation Group, Johns Hopkins University, the Joint Institute for Nuclear Astrophysics, the Kavli Institute for Particle Astrophysics and Cosmology, the Korean Scientist Group, the Chinese Academy of Sciences (LAMOST), Los Alamos National Laboratory, the Max-Planck-Institute for Astronomy (MPIA), the Max-Planck-Institute for Astrophysics (MPA), New Mexico State University, Ohio State University, University of Pittsburgh, University of Portsmouth, Princeton University, the United States Naval Observatory, and the University of Washington.

\end{document}